\documentstyle[amssymb,12pt,axodraw]{article}
\begin{document}

\begin{center}
{\Large {\bf Masses of Fermions in Supersymmetric Models}}\\

C.M. Maekawa and M. C. Rodriguez \\
{\it Funda\c c\~ao Universidade Federal do Rio Grande-FURG \\
Departamento de F\'\i sica \\
Av. It\'alia, km 8, Campus Carreiros \\
96201-900, Rio Grande, RS \\
Brazil}
\end{center}

\begin{abstract}
We consider the mass generation for the usual quarks and leptons in some
supersymmetric models. The masses of the top, the bottom, the charm, the tau
and the muon are given at the tree level. All the other quarks and the
electron get their masses at the one loop level in the Minimal Supersymmetric
Standard Model (MSSM) and in two Supersymmetric Left-Right Models, one model
uses triplets (SUSYLRT) to break $SU(2)_{R}$-symmetry and the other use doublets(SUSYLRD).

\end{abstract}

\section{Introduction}

One of important issues for particle physics is the small mixing angles in the
charged fermion sector and the hierarchy of quark and lepton masses
\cite{sg}.This issue has brought into the focus due to\ the success of the
Standard Model (SM) in describing the available experimental data except such
mass spectrum and due to the recent experiments are showing it is likely
neutrinos have such mass hierarchy but their mix pattern differs from that of
the quarks, e.g., the 2-3 lepton mixing angle is close to the maximal value
while the analogous quark mixing angle is small ($\theta_{23}^{q}\sim2%
{{}^\circ}%
$). The fermions mass spectrum is an aspect of a problem named as fermion
flavour structure \cite{Moha99} which includes the suppression of flavour
change neutral current, strong CP-problem, etc. 

Approaches based on Supersymmetry (SUSY) have been proposed in order to
explain the values of these masses and the Cabibbo-Kobayashi-Maskawa (CKM)
matrix elements. They are guided by the pattern of hierarchy and one pattern
used is the following \textit{horizontal hierarchy}:%

\begin{equation}
\begin{array}[c]{ccccc}%
m_{t}:m_{c}:m_{u} & \sim & 1:\varepsilon_{u}:\varepsilon_{u}^{2} &  &
\varepsilon_{u}\simeq\frac{1}{500},\\
m_{b}:m_{s}:m_{d} & \sim & 1:\varepsilon_{d}:\varepsilon_{d}^{2} &  &
\varepsilon_{d}\simeq\frac{1}{50},\\
m_{\tau}:m_{\mu}:m_{e} & \sim & 1:\varepsilon_{e}:\varepsilon_{e}^{2} &  &
\varepsilon_{e}\simeq\frac{1}{50},
\end{array}
\end{equation}
where $m_{u}$ and $m_{d}$ are current quark masses. This pattern has
suggested, e.g., the masses of different families are generated in different
stages of chiral symmetry breaking: at the first stage only $t$ and $b$ quarks
acquire mass and there is no mixing, at the second stage $c$ and $s$ get mass
and there is a mixing between the third and second family and in the end $u$
and $d$ quarks get their masses. This can be realized by the \textit{radiative
mass generation mechanism} where the lowest quarks are prevented to acquire
mass at tree level \cite{banks,ma,Pokors+90}. However this mechanism in
supersymmetric models gives rise to the flavour changing problem in the loop
that generates the masses. In order to avoid this problem \ a horizontal
flavour symmetry has been proposed within supersymmetric extensions of SM
\cite{Barbieri+96} and unified $SO\left(  10\right)  $ model \cite{Berez+05}. The
last one assumes a pattern where the first family instead of third
family plays a unique role and named it as \textit{inverse hierarchy pattern}.
 This is inspired by the fact at GUT scale running
masses of electron, u and d quarks are not strongly split.

Another pattern shows us two different scales for the masses of quarks, one is
at MeV scale
\begin{equation}
m_{u}\sim1-5\;MeV,\,\ m_{d}\sim3-9MeV,\;m_{s}\sim75-170MeV,
\end{equation}
while the other is at GeV scale:
\begin{equation}
m_{c}\sim1,15-1,35\;GeV,\,\ m_{t}\sim174\;GeV,\;m_{b}\sim4,0-4,4\;GeV.
\end{equation}

This point of view has implications for nuclear physics. Due to $u$, $d$ and
$s$ quarks are lights one is allowed to build an effective field theory as an
expansion on masses of light quarks of the underlying theory. The Chiral
Perturbation Theory (ChPT) \cite{CHPT99} is the prototype of this approach. It
respects all principles of the underlying theory but with effective degrees of
freedom instead of quarks degrees of freedom. A model independent description
of dynamics \cite{PotNN} and structure of nucleons \cite{FFnuc} above MeV
scale is obtained.

We explore the implications of this picture in MSSM
 and Left-Right Supersymmetric Model (LRSM). In the framework of
SUSY models the Higgs mechanism can be extended by increasing the number of
scalar particles, as a consequence the number of vacuum expectation values
also is increased and one has the possibility of two scale of masses for the
case of two scalar particles. However there is no constrains to the size of
masses and it is likely they could be at the same scale.

The mechanism of radiative mass generation implemented in SUSY models based on
the last pattern allows a small values for the mixing angles of
Cabibbo-Kobayashi-Maskawa (CKM) matrix since the beginning\ and two mass
scales once one requires all the three heavy quarks ($t$, $b$ and $c$ )
acquire mass at tree level, while an additional symmetry \cite{banks,ma}
suppresses the mass generation of light quarks ($u$, $d$ and $s$) . For the
leptons the same description is applied and a low value for the masses of
light quarks and light lepton and the Cabibbo-Kobayashi-Maskawa (CKM) matrix
in agreement with the experimental limits is obtained. The flavour change
problem is under control.

The outline of this work is as follows. The section \ref{sec:ma}
describes how the additional discrete symmetry ${\cal Z}^{\prime}_{2}$ is 
introduced into the framework of MSSM in order to prevent the light
quarks and the electron to acquire mass at tree level. The radiative mechanism
is described in section \ref{sec:soft}, and $u$, $d$
and $s$ quarks together with the electron acquire mass at 1-loop level. We also 
show that our results are still
valid in two supersymmetric left-right models. Our conclusions are found in
the last section. All the details of the models (conventions) and computations of mass
matrices are in the appendices.

\section{MSSM and ${\cal Z}^{\prime}_{2}$ Symmetry.}

\label{sec:ma}

In the MSSM \cite{mssm}, which the gauge group is $SU(3)_{C}\otimes
SU(2)_{L}\otimes U(1)_{Y},$ let $\hat{L}$ ($\hat{l}^{c}$) denotes left-handed
(right-handed) leptons \footnote{$c$ stands for charge conjugation}, $\hat{Q}$
$(\hat{u},\hat{d}^{c})$ left-handed (right-handed) quarks and $\hat{H}%
_{1},\hat{H}_{2}$ are the Higgs doublets respectively (a summary is in
appendix \ref{apend:mssm}).

The fermion mass comes from the following terms of the superpotential 
(Eq.(\ref{mssmrpc})):
\begin{equation}
W=-\left(  y_{ab}^{l}L_{a}H_{1}l_{b}^{c}+y_{ij}^{d}Q_{i}H_{1}d_{j}^{c}%
+y_{ij}^{u}Q_{i}H_{2}u_{j}^{c}+h.c.\right)  + \ldots,
\end{equation}
where $y_{ab}^{l}$, $y_{ij}^{d}$ and $y_{ij}^{u}$ are the yukawa couplings of
Higgs with leptons families, ``down" sector quarks and ``up" sector quarks
respectively and $\ldots$stands for other terms which we are not concerned here.
The family indices $a$ and $i$ run over $e, \mu , \tau$ and 
$1,2,3$, respectively.

Based on Eq.(\ref{vevscalarmssm}), we get the following non-diagonal mass
matrices $M_{ij}^{l,d,u}$ :
\begin{eqnarray}
M_{ij}^{u}  & =& \frac{y_{ij}^{u}}{\sqrt{2}}v_{2}(u_{i}u_{j}^{c}+h.c.),\nonumber\\
M_{ij}^{d}  & =& \frac{y_{ij}^{d}}{\sqrt{2}}v_{1}(d_{i}d_{j}^{c}+h.c.),\nonumber\\
M_{ab}^{l}  & =& \frac{y_{ab}^{l}}{\sqrt{2}}v_{1}(l_{a}l_{b}^{c}+h.c.).
\label{fermionmassmssm}
\end{eqnarray}
Where all the fermions fields are still Weyl spinors. The fields in the 
parenthesis define the basis to get the mass matrix. We can also rewrite 
all the equations above as 
\begin{equation}
M^\psi_{ij}=- \left( \bar{\psi}_{iL}m_{ij}\psi_{jR}+h.c. \right),
\end{equation}
where $\psi_{i}$ \footnote{$\psi_{i}$ indicate 
any charged fermion, the translation of two-component formalism into four-component formalism 
can be found in \cite{mssm,dress,wb}} is the Dirac spinor.

Therefore, the ``down" quark sector ($d,s$ and $b$ quarks) as well as the
$e,\mu$ and $\tau$ will have masses proportional to the vacuum expectation
value $v_{1}$, whereas the ``up" sector  will have masses proportional to
$v_{2}$. Note that the neutrinos remains massless due to lepton-number
conservation, but we know that neutrinos have masses. In order to give mass to
neutrinos one has to introduce $R$ -parity violating term $W_{RV}$ of
Eq.(\ref{mssmrpc}). We will focus our attention to the quark and lepton sector
and for the case of neutrinos the reader is invited to look at
Ref.~\cite{lepmass}.

Although the Higgs mechanism and SUSY allow two different scale of masses,
there is no underlying principle to keep them different from each other. The
fact that $m_{u},m_{d},m_{s}$ and $m_{e}$ are many orders of magnitude smaller
than the masses of others fermions may well be indicative of a radiative
mechanism at work for these masses as considered at \cite{banks,ma}.

The key feature of this kind of mechanism is to allow only the quarks $c,b,t,$
and the leptons $\mu$ and $\tau$ have Yukawa couplings to the Higgs bosons. It
means to prevent $u,d,s$ and $e$ from picking up tree-level masses, all one
needs to do at this stage is to impose the following ${\cal Z}^{\prime}_{2}$ 
symmetry on the Lagrangian
\begin{equation}
\widehat{d}_{2}^{c}\longrightarrow-\widehat{d}_{2}^{c},\,\ \widehat{d}_{3}%
^{c}\longrightarrow-\widehat{d}_{3}^{c},\,\ \widehat{u}_{3}^{c}\longrightarrow
-\widehat{u}_{3}^{c},\,\ \widehat{l}_{3}^{c}\longrightarrow-\widehat{l}%
_{3}^{c},
\label{z2def}
\end{equation}
the others superfields are even under this symmetry. On Ref.~\cite{ma}, only
the electron and the first  quark family don't pick up tree-level masses. 

After the diagonalization procedure of mass matrices of fermions (see appendix
\ref{apend:massdiag}), we can write
$M_{diag}= \mbox{diag}(m_f{_1},m_{f_{2}},m_{f_{3}})$  where
\begin{equation}
m_{f_{1}}=\frac{1}{2}\left(  t_{f}+r_{f}\right)  ,\ \ m_{f_{2}}=\frac{1}{2}\left(
t_{f}-r_{f}\right)  ,\ m_{f_{3}}=0,\label{autovalor2}
\end{equation}
with $t_{f}$, $r_{f}$ are given at Eq.(\ref{autovalor3}) and $f$ runs over fermions. 
 Taking $M_{diag}$ into account we can do 
  the following phenomenological identification:
\begin{equation}
m_{u_{1}}\equiv m_{t},\,\ m_{d_{1}}\equiv m_{b},\,\ m_{u_{2}}\equiv m_{c},\,\ m_{l_{1}}\equiv
m_{\tau},\ m_{l_{2}}\equiv m_{\mu}\,\ .
\end{equation} 
In order to fit the 
experimental data we  make the following choices into Eq.(\ref{autovalor2}):
\begin{eqnarray}
&&\hspace{2.8cm}t_{l}  = m_{\tau}+m_{\mu},\,\,\,\ t_{u}=m_{t}+m_{c},\,\ t_{d}=m_{b},
\,\nonumber\\
&&(y_{13}^{l})^{2}u_{l}-2y_{12}^{l}y_{13}^{l}v_{l}+y_{l}^{2}+(y_{12}^{l})^{2}z_{l}  
= \frac{1}{4}\left[  (m_{\tau}+m_{\mu})^{2}-(m_{\tau}-m_{\mu})^{2}\right]
\,\ ,\nonumber\\
&&(y_{13}^{u})^{2}u_{u}-2y_{12}^{u}y_{13}^{u}v_{u}+y_{u}^{2}+(y_{12}^{u})^{2}z_{u}  
= \frac{1}{4}\left[  (m_{t}+m_{c})^{2}-(m_{t}-m_{c})^{2}\right]  \,\ . \nonumber 
\end{eqnarray}
Thus the quarks $u, d, s$ and the electron come about be massless due to
${\cal Z}^{\prime}_{2}$ symmetry.  It means a discrete symmetry like
${\cal Z}^{\prime}_{2}$ is protecting the Chiral symmetry to be broken in
the $SU\left(  3\right)  _{F}$ sector.

Now, we want to show that
a consistent picture with the experimental values of the
Cabibbo-Kobayashi-Maskawa (CKM) matrix is obtained even in the presence of the ${\cal Z}^{\prime}_{2}$ symmetry. The CKM matrix comes from the fact
that the mass eigenstates of physical quarks are a mixture of different quarks
eigenstates of symmetry and for three generation of quarks one has:

\begin{equation}
V_{CKM}=\left(
\begin{array}
[c]{ccc}%
V_{ud} & V_{us} & V_{ub}\\
V_{cd} & V_{cs} & V_{cb}\\
V_{td} & V_{ts} & V_{tb}%
\end{array}
\right)
\end{equation}
where the matrix element $V_{ij}$ indicates the contribution of quark ($j$ )
to quark ($i$). The experimental values are \cite{pdg}:
\begin{equation}
\left(
\begin{array}
[c]{ccc}%
0.9739-0.9751 & 0.221-0.227 & 0.0029-0.0045\\
0.221-0.227 & 0.9730-0.9744 & 0.039-0.044\\
0.0048-0.014 & 0.037-0.043 & 0.9990-0.9992
\end{array}
\right)  \label{datackm}%
\end{equation}

As the quarks $t$ and $c$ get masses at tree-level their states can be  mixed
and we can write the  eigenvector of  ``up" quark sector 
\footnote{$(t,c,u)^{T}=(u_{1},u_{2},u_{3})^{T}E_{L}^{uT}$} as
\begin{equation}
E_{L}^{u}=\left(
\begin{array}
[c]{ccc}%
\cos\theta & \sin\theta & 0\\
-\sin\theta & \cos\theta & 0\\
0 & 0 & 1
\end{array}
\right)  .\label{autovetor1}
\end{equation}
For another hand, in the ``down" quark sector only the quark $b$ get mass at
tree-level and there is no mixing on this sector. Therefore we can write
\begin{equation}
E_{L}^{d}=I_{3 \times 3} 
\,\ .\label{autovetor2}
\end{equation}
where $I_{3 \times 3}$ is the identity matrix $3 \times 3$. Then, with 
Eq.(\ref{autovetor1},\ref{autovetor2}), we can get an expression to the
CKM matrix as follows:
\begin{equation}
V_{CKM}=E_{L}^{u\dagger}E_{L}^{d}=\left(
\begin{array}
[c]{ccc}%
\cos\theta & -\sin\theta & 0\\
\sin\theta & \cos\theta & 0\\
0 & 0 & 1
\end{array}
\right)  \,\ .\label{ckm}
\end{equation}

Comparing Eqs(\ref{datackm},\ref{ckm}), we can conclude that the
${\cal Z}^{\prime}_{2}$ symmetry in the MSSM can explain the lower masses
of the $u,d$ and $s$ quarks and also gives a hint about the 
mixing angles of quarks. 

\section{Radiative Mechanism to the fermions masses}

\label{sec:soft}

The discrete symmetry ${\cal Z}^{\prime}_{2}$ has to be broken in order 
to allow the generation of fermions masses by radiative corrections and  
the  most general soft supersymmetry breaking Lagrangian 
Eq.(\ref{iupa}) has already the following ${\cal Z}^{\prime}_{2}$ breaking
terms 
\begin{eqnarray}
{\cal L}_{soft}&=&\left[ \sum_{i=1}^{3}A_{i3}^{d}H_{1}\tilde{Q}_{i}
\tilde{d}_{3L}^{c}+\sum_{i=1}^{3}A_{i2}^{d}H_{1}\tilde{Q}_{i}
\tilde{d}_{2L}^{c}+\sum_{i=1}^{3}A_{i3}^{u}H_{2}\tilde{Q}_{i}
\tilde{u}_{3L}^{c} \right. \nonumber \\
&+& \left. \sum_{a=1}^{3}A_{a3}^{l}H_{1}\tilde{L}_{a}\tilde
{l}_{3L}^{c}+h.c. \right] + \ldots ,
\end{eqnarray}
where $\ldots$ stands for other terms. It means that the squarks $\tilde{q}$ and
$\tilde{q}^{c}$ will mix, the same will happen with the sleptons $\tilde{l}$
and $\tilde{l}^{c}$ \footnote{In supersymmetric theories the sfermions masses come 
from the scalar potetial given by $V=V_{F}+V_{D}+V_{soft}$ \cite{dress,wb}, here we 
are not showing all the details}. We can write
\begin{eqnarray}
{\cal L}_{soft}  & =& M_{Q}^{2}\left(  \tilde{u}_{3}^{\ast}\tilde{u}_{3}+\tilde{d}%
_{3}^{\ast}\tilde{d}_{3}+\tilde{d}_{2}^{\ast}\tilde{d}_{2}\right)  +M_{u}
^{2}\tilde{u}_{3}^{c\ast}\tilde{u}_{3}^{c}+M_{d}^{2}\left(  \tilde{d}%
_{3}^{c\ast}\tilde{d}_{3}^{c}+\tilde{d}_{2}^{c\ast}\tilde{d}_{2}^{c}\right)
+M_{L}^{2}\tilde{l}_{3}^{\ast}\tilde{l}_{3}\nonumber\\
& +&M_{l}^{2}\tilde{l}_{3}^{c\ast}\tilde{l}_{3}^{c}+\left[  A_{33}^{l}%
v_{1}\tilde{l}_{3}\tilde{l}_{3}^{c}+A_{33}^{u}v_{2}\tilde{u}_{3}\tilde{u}%
_{3}^{c}+A_{33}^{d}v_{1}\tilde{d}_{3}\tilde{d}_{3}^{c}+A_{22}^{d}v_{1}%
\tilde{d}_{2}\tilde{d}_{2}^{c}+h.c.\right]  \nonumber \\ 
&+& \ldots \label{misturasquarks}
\end{eqnarray}
For the case of the physical u-squark states, that we will denotated as $\tilde{u}_{1},\tilde{u}_{2}$, 
it gives rise to the following eigenstates of mass as functions of symmetry eigenstates
\footnote{Mixing between squarks of different generations can cause severe problems due to too 
large loop contributions to flavour changing neutral currents (FCNC) process. Due this fact we 
are ignoring intergenerational mixing.}:
\begin{eqnarray}
\tilde{u}_{1}  & =& \cos\theta_{\tilde{u}}\tilde{u}_{3}+
\sin\theta_{\tilde{u}}\tilde{u}^{c\ast}_{3},\nonumber\\
\tilde{u}_{2}  & =&- \sin\theta_{\tilde{u}}\tilde{u}_{3}+
\cos\theta_{\tilde{u}}\tilde{u}^{c\ast}_{3}.
\end{eqnarray}
Similar expressions to the  d-squark ($\tilde{d}$), s-squark ($\tilde{s}$),  
 selectron ($\tilde{e}$) and smuon ($\tilde{ \mu}$) can be obtained.  
By another side, the interaction between quark-squark-gluino is given
by:
\begin{eqnarray}
{\cal L}_{q \tilde{q}\tilde{g}}  & =-i\sqrt{2}g_{s}\bar{T}^{a}(
\tilde{u}^{c}_{i}\bar{u}^{c}_{i}
\bar{\lambda}_{C}^{a}-\bar{\tilde{u}}^{c}_{i}u^{c}_{i}\lambda_{C}^{a}+
\tilde{d}^{c}_{i}\bar{d}^{c}_{i}\bar{\lambda}_{C}^{a}-
\bar{\tilde{d}}^{c}_{i}d^{c}_{i}\lambda_{C}^{a})+ \ldots \nonumber\label{vertmssm}\\
\end{eqnarray}
This interaction generate the radiative mechanism for the mass of the $u,d$
and $s$ quarks. On Fig.(1) we show the lowest order contribution. It was also
shown in \cite{banks,ma} to current mass of up quark. 
\begin{figure}[t]
\begin{center}
\begin{picture}(300,56)(0,0)
\Vertex(180,10){1.5}
\Text(180,5)[l]{$g_{s}$}
\Vertex(120,10){1.5}
\Text(120,5)[r]{$g_{s}$}
\ArrowLine(100,10)(200,10)
\Text(80,10)[l]{$u_{L}$}
\Text(205,10)[l]{$u^{c}_{L}$}
\DashCArc(150,10)(30,0,180){4}
\Text(150,42)[c]{$\otimes$}
\Text(150,50)[c]{$m^{2}_{\tilde{u}}$}
\Text(150,12)[c]{$X$}
\Text(150,20)[c]{$m_{\tilde{g}}$}
\Text(130,3)[l]{$\tilde{g}^{a}$}
\Text(160,3)[l]{$\bar{\tilde{g}}^{a}$}
\Text(120,30)[r]{$\tilde{u}_{\alpha}$}
\Text(180,30)[l]{$\tilde{u}_{\alpha}^{c}$}
\end{picture}\\[0pt]\textsl{Figure 1: The diagram which gives mass to quark
$u$ which does not apperar in the superpotential, $\tilde{g}$ is the gluino
while $\tilde{u}$ is the squark and $\alpha=1,2$.}
\end{center}
\par
\label{figone}\end{figure}

Notice that, all the mass insertion on this diagram
came from the soft term, see Eq.(\ref{misturasquarks}), while the two vertices come from
Eq.(\ref{vertmssm}). Similar diagram can be drawn to the $d$ and $s$ quarks.

Following \cite{ma} we calculated their masses and we obtained:
\begin{eqnarray}
m_{u}  & \propto& \frac{\alpha_{s}\sin(2\theta_{\tilde{u}})}{\pi}m_{\tilde{g}%
}\left[  \frac{M_{\tilde{u_{1}}}^{2}}{M_{\tilde{u_{1}}}^{2}-m_{\tilde{g}}^{2}%
}\ln\left(  \frac{M_{\tilde{u_{1}}}^{2}}{m_{\tilde{g}}^{2}}\right)  \right.
\nonumber\\
& -& \left.  \frac{M_{\tilde{u_{2}}}^{2}}{M_{\tilde{u_{2}}}^{2}-m_{\tilde{g}%
}^{2}}\ln\left(  \frac{M_{\tilde{u_{2}}}^{2}}{m_{\tilde{g}}^{2}}\right)
\right]  \,\ ,\nonumber\\
m_{d}  & \propto& \frac{\alpha_{s}\sin(2\theta_{\tilde{d}})}{\pi}m_{\tilde{g}%
}\left[  \frac{M_{\tilde{d_{1}}}^{2}}{M_{\tilde{d_{1}}}^{2}-m_{\tilde{g}}^{2}%
}\ln\left(  \frac{M_{\tilde{d_{1}}}^{2}}{m_{\tilde{g}}^{2}}\right)  \right.
\nonumber\\
& -& \left.  \frac{M_{\tilde{d_{2}}}^{2}}{M_{\tilde{d_{2}}}^{2}-m_{\tilde{g}%
}^{2}}\ln\left(  \frac{M_{\tilde{d_{2}}}^{2}}{m_{\tilde{g}}^{2}}\right)
\right]  \,\ ,\nonumber\\
m_{s}  & \propto& \frac{\alpha_{s}\sin(2\theta_{\tilde{s}})}{\pi}m_{\tilde{g}%
}\left[  \frac{M_{\tilde{s_{1}}}^{2}}{M_{\tilde{s_{1}}}^{2}-m_{\tilde{g}}^{2}%
}\ln\left(  \frac{M_{\tilde{s_{1}}}^{2}}{m_{\tilde{g}}^{2}}\right)  \right.
\nonumber\\
& -& \left.  \frac{M_{\tilde{s_{2}}}^{2}}{M_{\tilde{s_{2}}}^{2}-m_{\tilde{g}%
}^{2}}\ln\left(  \frac{M_{\tilde{s_{2}}}^{2}}{m_{\tilde{g}}^{2}}\right)
\right]  \,\ ,
\end{eqnarray}
where $m_{\tilde{g}}$ , $m_{\tilde{u}}$, $m_{\tilde{d}}$ and$,m_{\tilde{s}%
}^{2}$ are, respectively, the masses of the gluino, u-squark,  d-squark and
s-squark.

In order to obtain quark masses in agreement with the experimental limits
\cite{pdg} we set $m_{\tilde{g}}\approx100$ GeV,  $\sin(2\theta_{\tilde{u}%
})\approx\sin(2\theta_{\tilde{d}})\approx10^{-3}$ and $\sin(2\theta_{\tilde
{s}})\approx10^{-2}$.

The electron couples with the gaugino $\lambda_{B}$ of $U(1)$ group as the
following:
\begin{equation}
{\cal L}_{l \tilde{l}\tilde{g}}=-\frac{ig^{\prime}}{\sqrt{2}}(2)\left(  
\tilde{l}^{c}_{a}\bar{l}^{c}_{a}\bar{\lambda}_{B}-
\bar{\tilde{l}}^{c}_{a}l^{c}_{a}\lambda_{B}\right) + \ldots  .
\end{equation}
This allows the diagram of Fig.(2) to contribute to the electon mass.
Therefore the electron mass is given by:
\begin{figure}[t]
\begin{center}
\begin{picture}(300,56)(0,0)
\Vertex(180,10){1.5}
\Text(180,5)[l]{$g^{\prime}$}
\Vertex(120,10){1.5}
\Text(120,5)[r]{$g^{\prime}$}
\ArrowLine(100,10)(200,10)
\Text(80,10)[l]{$e_{L}$}
\Text(205,10)[l]{$e^{c}_{L}$}
\DashCArc(150,10)(30,0,180){4}
\Text(150,42)[c]{$\otimes$}
\Text(150,50)[c]{$m^{2}_{\tilde{e}}$}
\Text(150,12)[c]{$X$}
\Text(150,20)[c]{$m^{\prime}$}
\Text(130,3)[l]{$\lambda_{B}$}
\Text(160,3)[l]{$\bar{\lambda}_{B}$}
\Text(120,30)[r]{$\tilde{e}_{\alpha}$}
\Text(180,30)[l]{$\tilde{e}_{\alpha}^{c}$}
\end{picture}\\[0pt]\textsl{Figure 2: The diagram which gives mass to electron
which does not apperar in the superpotential, $\tilde{e}$ is the selectron and
$\alpha=1,2$.}
\end{center}
\par
\label{figone}\end{figure}

\begin{eqnarray}
m_{e}  & \propto& \frac{\alpha_{U(1)}\sin(2\theta_{\tilde{e}})}{\pi}m^{\prime
}\left[  \frac{M_{\tilde{e_{1}}}^{2}}{M_{\tilde{e_{1}}}^{2}-m^{\prime2}}%
\ln\left(  \frac{M_{\tilde{e_{1}}}^{2}}{m^{\prime2}}\right)  \right.
\nonumber\\
& -& \left.  \frac{M_{\tilde{e_{2}}}^{2}}{M_{\tilde{e_{2}}}^{2}-m^{\prime2}}%
\ln\left(  \frac{M_{\tilde{e_{2}}}^{2}}{m^{\prime2}}\right)  \right]  \,\ ,
\end{eqnarray}
where $\alpha_{U(1)}=g^{\prime2}/(4\pi)$. Similar numerical analysis can be
done as we performed in the quark sector above.

From the Figs.(1,2) one can see why  quarks are heavier than
 leptons; they get color contribution while leptons not as was
showed on Ref.~\cite{banks}.

\section{Supersymmetric Left-Right Model (SUSYLR)}

\label{sec:susylr}

The supersymmetric extension of left-right models \cite{susylr,doublet} is
based on the gauge group $SU(3)_{C}\otimes SU(2)_{L}\otimes SU(2)_{R}\otimes
U(1)_{B-L}$. Apart from its original motivation of providing a dynamic
explanation for the parity violation observed in low-energy weak interactions,
this model differs from the SM in another important aspect; it explains the
observed lightness of neutrinos in a natural way and it can also solve the
strong CP problem.

On the technical side, the left-right symmetric model has a problem similar to
that in the SM: the masses of the fundamental Higgs scalars diverge
quadratically. As in the SM, the SUSYLR can be used to stabilize the scalar
masses and cure this hierarchy problem. SUSYLR models have the additional
appealing characteristics of having automatic R-parity conservation.

On the literature there are two different SUSYLR models. They differ in their
$SU(2)_{R}$ breaking fields: one uses $SU(2)_{R}$ triplets (SUSYLRT) and the
other $SU(2)_{R}$ doublets (SUSYLRD). Theoretical consequences of these models
can be found in various papers including \cite{susylr} and \cite{doublet}
respectively. Some details of both models are described at
Appendix~\ref{apend:susylrt} and Appendix~\ref{apend:susylrd}.

\section{Masses of Fermions in SUSYLR Models}

For SUSYLRT, the mass term to the quarks is ( Eqs.\ref{suplr},\ref{suplrd}):
\begin{equation}
{\cal L}^{mass}_{quarks}=-\left[  h_{ij}^{q}Q_{i}^{T}\Phi\imath\tau_{2}Q_{j}^{c}+\tilde{h}_{ij}%
^{q}Q_{i}^{T}\Phi^{\prime}\imath\tau_{2}Q_{j}^{c}+h.c.\right]  .
\end{equation}
Using Eq.(\ref{vevsusylr}) on the equation above, we get the following mass matrix in the
non-diagonal form
\begin{eqnarray}
M_{ij}^{u}  & =\frac{1}{\sqrt{2}}\left[  k_{1}h_{ij}^{q}+k_{2}^{\prime}%
\tilde{h}_{ij}^{q}\right]  (u_{i}u_{j}^{c}+hc),\nonumber\\
M_{ij}^{d}  & =\frac{1}{\sqrt{2}}\left[  k_{1}^{\prime}h_{ij}^{q}+k_{2}%
\tilde{h}_{ij}^{q}\right]  (d_{i}d_{j}^{c}+hc).\label{quarkmasssusylr}%
\end{eqnarray}

For the leptons (Eq.\ref{suplr}), the mass term is
\begin{eqnarray}
{\cal L}^{mass}_{leptons}& =&-\left[  f_{ab}(L_{a}^{T}\imath\tau_{2}\Delta_{L}L_{b})+
f_{ab}^{c}(L_{a}^{cT}\imath\tau_{2}\delta_{L}^{c}L_{b}^{c})+h_{ab}^{l}(L_{a}^{T}
\Phi\imath\tau_{2}L_{b}^{c})\right. \nonumber \\
&+& \left. \tilde{h}_{ab}^{l}(L_{a}^{T}\Phi^{\prime}\imath\tau_{2}L_{b}^{c}) + 
h.c.\right] \,\ .
\end{eqnarray}
Using Eq.(\ref{vevsusylr}) on equation given above, we get the following mass matrix in the
non-diagonal representation
\begin{eqnarray}
M_{ab}^{l}  & =&\frac{1}{\sqrt{2}}\left[  k_{1}^{\prime}h_{ab}^{l}+
k_{2}\tilde{h}_{ab}^{l}\right]  (l_{a}l_{b}^{c}+hc),\nonumber\label{lepmass1}\\
M_{ab}^{\nu}  & =&\frac{1}{\sqrt{2}}\left[  k_{1}h_{ab}^{l}+k_{2}^{\prime
}\tilde{h}_{ab}^{l}\right]  (\nu_{a}\nu_{b}^{c}+hc)+
\frac{\upsilon_{R}}{\sqrt{2}}f_{ab}^{c}(\nu_{a}^{c}\nu_{b}^{c}+hc)\nonumber\\
& -&\frac{\upsilon_{L}}{\sqrt{2}}f_{ab}(\nu_{a}\nu_{b} +hc).\label{lepmasssusylr}
\end{eqnarray}
This result is in agreement with the presented in \cite{mfrank}, if we take
$\upsilon_{L}=0$.

For another hand,  in the case of SUSYLRD one extracts from Eqs.(\ref{suplrd})
the mass term to the leptons
\begin{equation}
{\cal L}^{mass}_{leptons}=-\left(  h_{ab}^{l}(L_{a}^{T}\Phi\imath\tau_{2}L_{b}^{c})+\tilde{h}%
_{ab}^{l}(L_{a}^{T}\Phi^{\prime}\imath\tau_{2}L_{b}^{c})+h.c.\right) \,\ .
\end{equation}
Using Eq.(\ref{vevsusylr}) above, we get the following mass matrix in the non
diagonal representation
\begin{eqnarray}
M_{ab}^{l}  & =-\frac{1}{\sqrt{2}}\left[  k_{1}^{\prime}h_{ab}^{l}+k_{2}%
\tilde{h}_{ab}^{l}\right]  (l_{a}l_{b}^{c}+hc),\nonumber\\
M_{ab}^{\nu}  & =-\frac{1}{\sqrt{2}}\left[  k_{1}h_{ab}^{l}+k_{2}^{\prime
}\tilde{h}_{ab}^{l}\right]  (\nu_{a}\nu_{b}^{c}+hc).\label{lepmass2}%
\end{eqnarray}
From Eqs.(\ref{lepmass1}, \ref{lepmass2}) we see that the choice of the
triplets is preferable to doublets because in the first case we can
generate a large Majorana mass for the right-handed neutrinos \cite{mfrank}.

The $d,s$ and $b$ quarks as well as the $e,\mu$ and $\tau$ leptons will have
masses proportional to the vacuum expectation values $k_{1}^{\prime},k_{2}$,
whereas the $u,c$ and $t$ will have masses proportional to $k_{1}%
,k_{2}^{\prime}$.

Now, we are deal with the charged fermions and we are going to present the results which are 
hold in both models. To avoid flavor-changing-neutral
currents (see \cite{simp}), we can choose the vacuum
expectations values of the bidoublets as
\begin{equation}
\left\langle \Phi\right\rangle =\frac{1}{\sqrt{2}}\left(
\begin{array}
[c]{cc}%
k_{1} & 0\\
0 & 0
\end{array}
\right)  ;\,\ \left\langle \Phi^{\prime}\right\rangle =\frac{1}{\sqrt{2}%
}\left(
\begin{array}
[c]{cc}%
0 & 0\\
0 & k_{2}%
\end{array}
\right)  .
\end{equation}
Where $k_{1},k_{2}$ are of the order of the electroweak scale $10^{2}$GeV.
Using this fact on Eqs.(\ref{quarkmasssusylr},\ref{lepmasssusylr}), we can
rewrite the mass matrix of charged fermion as
\begin{eqnarray}
M_{ij}^{u}  & =\frac{h_{ij}^{q}}{\sqrt{2}}k_{1}(u_{i}u_{j}^{c}+hc),\nonumber\\
M_{ij}^{d}  & =\frac{\tilde{h}_{ij}^{q}}{\sqrt{2}}k_{2}(d_{i}d_{j}%
^{c}+hc),\nonumber\\
M_{ab}^{l}  & =-\frac{\tilde{h}_{ab}^{l}}{\sqrt{2}}k_{2}(l_{i}l_{j}%
^{c}+hc).\label{quarkmasssusylrsimp}%
\end{eqnarray}
The equations above are very similar to ones we get on the MSSM case, see
Eq.(\ref{fermionmassmssm}). Following the references \cite{ma,banks} we can
try to find a discrete symmetry in order to prevent the electron and the quarks $u$ and
$d$ from acquire masses at tree level. We impose the following 
${\cal Z}^{\prime}_{2}$ symmetry
\begin{eqnarray}
\hat{Q}_{2}^{c}  & \rightarrow& \tau_{3}\,\ \hat{Q}_{2}^{c},\nonumber\\
\hat{Q}_{3}^{c}  & \rightarrow& -I\,\ \hat{Q}_{3}^{c},\nonumber\\
\hat{L}_{3}  & \rightarrow& \tau_{3}\,\ \hat{L}_{3},\label{z2susylr}
\end{eqnarray}
and the others superfields are even under this symmetry (compare with
Eq.(\ref{z2def})). With Eqs.(\ref{z2susylr},\ref{mixingquarkssusylr}) at hand we can
reproduce the results presented in the section \ref{sec:ma}. We want to
emphasize  this symmetry does not forbid a large Majorana mass for the
right-handed neutrinos and the results presented in Ref.~\cite{mfrank} are
still valid.

The mixing between the squarks and the sleptons is given by
\begin{eqnarray}
{\cal L}& =&m_{Q_{L}}^{2}(\tilde{u}_{3}^{\ast}\tilde{u}_{3}+\tilde
{d}_{3}^{\ast}\tilde{d}_{3})+m_{Q_{R}}^{2}(\tilde{u}_{3}^{\ast c}\tilde{u}%
_{3}^{\ast c}+\tilde{d}_{3}^{\ast c}\tilde{d}_{3})+m_{L_{L}}^{2}\tilde{l}%
_{3}^{\ast}\tilde{l}_{3}+m_{L_{R}}^{2}\tilde{l}_{3}^{c}\tilde{l}_{3}^{\ast
c}+\nonumber\\
& +&\frac{1}{\sqrt{2}}\left[  
A^{LR}_{33}k_{1}\tilde{l}_{3}\tilde{l}_{3}^{c}
+\tilde{A}^{LR}_{33}k_{2}^{\prime}\tilde{l}_{3}\tilde{l}_{3}^{c}+
A^{QQ}_{33}k_{1}\tilde{u}_{3}\tilde{u}_{3}^{c}+
\tilde{A}^{QQ}_{33}k_{2}^{\prime}\tilde{d}_{3}\tilde{d}_{3}^{c}\right] \,\ , 
\label{mixingquarkssusylr}
\end{eqnarray}
which generates the diagrams shown in Figs.(1,2), and then we can reproduce
the results presented in the section \ref{sec:soft}.

\section{\textbf{Conclusions}}

\label{sec:concl}

We show that we can introduce a discrete symmetry ${\cal Z}^{\prime}_{2}$
in MSSM and in both SUSYLR in order to explain the lower masses of the quarks
$u,d$ and $s$ and of the electron while a consistent picture with experimental data of
CKM matrix is obtained. We have also shown that in the models studied in this work 
the heavy leptons ($\mu$ and $\tau$) acquire mass at tree level while the electron get
their mass at 1-loop level. 

  A discrete symmetry like ${\cal Z}^{\prime}_{2}$ protects
the Chiral symmetry to be broken in $SU\left(  3\right)  $ sector. This allows Chiral symmetry to be broken at different scales and two scales of mass is obtained. 

These results presented in this article is easily
extended to the others supersymmetric models as of the Ref.~\cite{susy331}.


\begin{center}
{\bf Acknowledgment}
\end{center}

C.M. Maekawa was supported by FAPERGS PROADE-2 under contract number 02/1266-6.

\appendix

\section{MSSM}

\label{apend:mssm}

\begin{table}[t]
\center
\renewcommand{\arraystretch}{1.5}
\begin{tabular}
[c]{|l|cc|cc|}\hline
Superfield & Usual Particle & Spin & Superpartner & Spin\\\hline\hline
\quad$\hat{V}^{\prime}$ (U(1)) & $V_{m}$ & 1 & $\lambda_{B}\,\,$ & $\frac
{1}{2}$\\
\quad$\hat{V}^{i}$ (SU(2)) & $V^{i}_{m}$ & 1 & $\lambda^{i}_{A}$ & $\frac
{1}{2}$\\
\quad$\hat{V}^{a}_{c} (SU(3))$ & $G^{a}_{m}$ & 1 & $\tilde{ g}^{a}$ &
$\frac{1}{2}$\\\hline
\quad$\hat{Q}_{i}\sim({\bf 3},{\bf 2},1/3)$ & $(u_{i},\,d_{i})_{L}$ & $\frac
{1}{2}$ & $(\tilde{ u}_{iL},\,\tilde{ d}_{iL})$ & 0\\
\quad$\hat{u}^{c}_{i}\sim({\bf 3^{\ast}},{\bf 1},-4/3)$ & $\bar{u}^{c}_{iL}$ &
$\frac{1}{2}$ & $\tilde{ u}^{c}_{iL}$ & 0\\
\quad$\hat{d}^{c}_{i}\sim({\bf 3^{\ast}},{\bf 1},2/3))$ & $\bar{d}^{c}_{iL}$ &
$\frac{1}{2}$ & $\tilde{ d}^{c}_{iL}$ & 0\\\hline
\quad$\hat{L}_{a}\sim({\bf 1},{\bf 2},-1)$ & $(\nu_{a},\,l_{a})_{L}$ & $\frac
{1}{2}$ & $(\tilde{ \nu}_{aL},\,\tilde{ l}_{aL})$ & 0\\
\quad$\hat{l}^{c}_{a}\sim({\bf 1},{\bf 1},2)$ & $\bar{l}^{c}_{aL}$ & $\frac{1}%
{2}$ & $\tilde{ l}^{c}_{aL}$ & 0\\\hline
\quad$\hat{H}_{1}\sim({\bf 1},{\bf 2},-1)$ & $(H_{1}^{0},\, H_{1}^{-})$ & 0 &
$(\tilde{ H}_{1}^{0},\, \tilde{ H}_{1}^{-})$ & $\frac{1}{2}$\\
\quad$\hat{H}_{2}\sim({\bf 1},{\bf 2},1)$ & $(H_{2}^{+},\, H_{2}^{0})$ & 0 &
$(\tilde{ H}_{2}^{+},\, \tilde{ H}_{2}^{0})$ & $\frac{1}{2}$\\\hline
\end{tabular}
\renewcommand{\arraystretch}{1}\caption{Particle content of MSSM.}%
\label{tab:mssm}%
\end{table}This model contains the particle content given at
Tab.(\ref{tab:mssm}). The  families index 
for leptons are $a,b$ and they run over
$e,\mu,\tau$, while the families index for the quarks are $i,j=1,2,3$. The 
parentheses in the first column  are the transformation properties under the respective representation of $(SU(3)_C,SU(2)_L,U(1)_Y)$.

The superfields formalism is useful in writing the manifestly invariant supersymmetric Lagrangian \cite{wb}. 
The fermions and scalars are represented by chiral superfields while the gauge bosons by
vector superfields. As usual the superfield of a field $\phi$ is denoted
by $\hat{\phi}$~\cite{mssm}. The chiral superfield of a multiplet $\phi$ is
denoted by
\begin{eqnarray}
\hat{\phi}\equiv\hat{\phi}(x,\theta,\bar{\theta})  & =&\tilde{\phi
}(x)+i\;\theta\sigma^{m}\bar{\theta}\;\partial_{m}\tilde{\phi}(x)+\frac{1}%
{4}\;\theta\theta\;\bar{\theta}\bar{\theta}\;\Box\tilde{\phi}(x)\nonumber\\
&+& \sqrt{2}\;\theta\phi(x)+\frac{i}{\sqrt{2}}\;\theta\theta
\;\bar{\theta}\bar{\sigma}^{m}\partial_{m}\phi(x)\nonumber\\
&+& \theta\theta\;F_{\phi}(x),\label{phi}%
\end{eqnarray}
while the vector superfield is given by
\begin{eqnarray}
\hat{V}\equiv\hat{V}(x,\theta,\bar{\theta})  & =&-\theta\sigma^{m}\bar{\theta
}V_{m}(x)+i\theta\theta\bar{\theta}\overline{\tilde{V}}(x)-i\bar{\theta}%
\bar{\theta}\theta\tilde{V}(x)\nonumber\\
& +& \frac{1}{2}\theta\theta\bar{\theta}\bar{\theta}D_{V}(x).\label{vector}%
\end{eqnarray}
The fields $F$ and $D$ are auxiliary fields which are needed to close the
supersymmetric algebra and eventually will be eliminated using their equations
of motion.

The Lagrangian of this model is written as
\begin{equation}
{\cal L}_{MSSM}={\cal L}_{SUSY}+{\cal L}_{soft}\,\ ,
\end{equation}
where $\mathcal{L}_{SUSY}$ is the supersymmetric piece and can be divided as
follows
\begin{equation}
{\cal L}_{SUSY}={\cal L}_{lepton}+{\cal L}_{Quarks}+{\cal L}_{Gauge}+
{\cal L}_{Higgs},
\end{equation}
where each term is given by 
\begin{eqnarray}
{\cal L}_{lepton}  & =& \int d^{4}\theta\;\left[  \,\hat{\bar{L}}_{a}e^{2g\hat
{V}+g^{\prime}\left(  -\frac{1}{2}\right)  \hat{V}^{\prime}}\hat{L}_{a}+\hat
{\bar{l^{c}}}_{a}e^{g^{\prime}\hat{V}^{\prime}}\hat{l}^{c}_{a}\,\right]
\,\ ,\nonumber\label{The Supersymmetric Term prop 2}\\
{\cal L}_{Quarks}  & =& \int d^{4}\theta\;\left[  \,\hat{\bar{Q}}_{i}
e^{2g_{s}\hat{V}_{c}^{a}+2g\hat{V}+g^{\prime}\left(  \frac{1}{6}\right)
\hat{V}^{\prime}}\hat{Q}_{i}+\hat{\bar{u}^{c}}_{i}e^{2g_{s}\hat{V}_{c}^{a}+g^{\prime
}\left(  -\frac{2}{2}\right)  \hat{V}^{\prime}}\hat{u}^{c}_{i}\right.
\nonumber\\
& +&\left.  \hat{\bar{d^{c}}}_{i}e^{2g_{s}\hat{V}_{c}^{a}+g^{\prime}\left(
\frac{1}{3}\right)  \hat{V}^{\prime}}\hat{d}^{c}_{i}\,\right]  \,\ ,\nonumber\\
{\cal L}_{Gauge}  & =& \frac{1}{4} \left \{ \int d^{2}\theta\;\left[  \sum_{a=1}%
^{8}W_{s}^{a\alpha}W_{s\alpha}^{a}+\sum_{i=1}^{3}W^{i\alpha}W_{\alpha}%
^{i}+W^{\prime\alpha}W_{\alpha}^{\prime} \right]\,+h.c.\right\}
\,, \nonumber \\
\label{The Supersymmetric Term prop 3}
\end{eqnarray}
the subscripts $a$ and $i$ are family index, summed over $e, \mu , \tau$ and 
$1,2,3$, respectively, on repetition. The last piece of our Lagrangian is written as
\begin{eqnarray}
{\cal L}_{Higgs}  & =& \int d^{4}\theta\;\left[  \,\hat{\bar{H}}_{1}%
e^{2g\hat{V}+g^{\prime}\left(  -\frac{1}{2}\right)  \hat{V}^{\prime}}\hat
{H}_{1}+\hat{\bar{H}}_{2}e^{2g\hat{V}+g^{\prime}\left(  \frac{1}{2}\right)
\hat{V}^{\prime}}\hat{H}_{2}\right]  \nonumber\\
& +& \int d^{2}\theta \,\ W+\int d^{2}\bar{\theta} \,\ \bar{W}
.\label{The Supersymmetric Term prop 4}%
\end{eqnarray}
The field strength are given by \cite{wb}
\begin{eqnarray}
W_{s\alpha}^{a}  & =&-\frac{1}{8g_{s}}\,\bar{D}\bar{D}e^{-2g_{s}\hat{V}_{c}%
^{a}}D_{\alpha}e^{2g_{s}\hat{V}_{c}^{a}}\,\ \alpha=1,2\,\ ,\nonumber\\
W_{\alpha}^{i}  & =&-\frac{1}{8g}\,\bar{D}\bar{D}e^{-2g\hat{V}^{i}}D_{\alpha
}e^{2g\hat{V}^{i}}\,\ ,\nonumber\\
W_{\alpha}^{\prime}  & =&-\frac{1}{4}\,DD\bar{D}_{\alpha}\hat{V}^{\prime
}\,\ .\label{fieldstrength}%
\end{eqnarray}

The superpotential is given by
\begin{eqnarray}
W  & =& W_{RC}+\bar{W}_{RC}+W_{RV}+\bar{W}_{RV},\nonumber\\
W_{2RC}  & =& \mu\epsilon\hat{H}_{1}\hat{H}_{2}+
y_{ab}^{l}\epsilon\hat{L}_{a}\hat{H}_{1}\hat{l}_{b}^{c}+y_{ij}^{u}\epsilon\hat{Q}_{i}\hat{H}_{2}
\hat{u}_{j}^{c}+y_{ij}^{d}\epsilon\hat{Q}_{i}\hat{H}_{1}\hat{d}_{j}^{c}%
,\nonumber\\
W_{RV}  & =& \mu_{1a}\epsilon \hat{L}_{a}\hat{H}_{2}+
\lambda_{abc}\epsilon \hat{L}_{a}\hat{L}_{b}\hat{l}_{c}^{c}
+\lambda^{1}_{aij}\epsilon \hat{L}_{a}\hat{Q}_{i}\hat{d}_{j}^{c}+
\lambda^{2}_{ijk}\hat{u}_{i}^{c}\hat{d}_{j}^{c}\hat{d}_{k}^{c}
\,\ .\label{mssmrpc}%
\end{eqnarray}
Where $W_{RC}$ ($W_{RV}$) conserves (violates) $R$-parity. The indices are summed on 
repetition.

The terms that break Supersymmetry softly and do not induce quadratic
divergence \cite{10} are
\begin{eqnarray}
\mathcal{L}_{\mathrm{soft}}  & =&-\frac{1}{2}\left(  \sum_{i=1}^{8}m_{\tilde
{g}}\tilde{g}^{i}\tilde{g}^{i}+\sum_{p=1}^{3}m_{\lambda}\lambda_{A}^{p}%
\lambda_{A}^{p}+m^{\prime}\lambda_{B}\lambda_{B}+h.c.\right)  -M_{L}^{2}%
\tilde{L}^{\dagger}\tilde{L} \nonumber \\
&-&M_{l}^{2}\tilde{l^{c}}^{\dagger}\tilde{l^{c}} 
-M_{Q}^{2}\tilde{Q}^{\dagger}\tilde{Q}-M_{u}^{2}\tilde{u^{c}}^{\dagger
}\tilde{u^{c}}-M_{d}^{2}\tilde{d^{c}}^{\dagger}\tilde{d^{c}}-M_{1}^{2}%
\tilde{H_{1}}^{\dagger}\tilde{H_{1}}-M_{2}^{2}\tilde{H_{2}}^{\dagger}%
\tilde{H_{2}} \nonumber \\
&-&\left[  A^{l}H_{1}\tilde{L}\tilde{l^{c}}
+  A^{u}H_{2}\tilde{Q}\tilde{u^{c}}+
A^{d}H_{1}\tilde{Q}\tilde{d^{c}}+
M_{12}^{2}H_{1}H_{2}+h.c.\right]  \,\ .
\label{iupa}
\end{eqnarray}
We have omitted generation indices and we do the same to all the soft terms 
that we will write on this article. The parameters  
$m_{\tilde{g}},m_{\lambda}$, and $m^{\prime}$ are the $SU(3),SU(2)$, and
$U(1)$ gaugino masses, respectively. $M_{1}^{2},M_{2}^{2}$ and $M_{12}^{2}$
are mass terms for the Higgs fields. The scalar mass terms $M_{L}^{2}%
,M_{l}^{2},M_{Q}^{2},M_{u}^{2}$, and $M_{d}^{2}$ are in general Hermitian
$3\times3$ matrices.

The vacuum expectation value of this model is given by
\begin{equation}
\left\langle H_{1}\right\rangle =\frac{1}{\sqrt{2}}\left(
\begin{array}
[c]{c}%
v_{1}\\
0
\end{array}
\right)  \,\ ,\,\ \left\langle H_{2}\right\rangle =\frac{1}{\sqrt{2}}\left(
\begin{array}
[c]{c}%
0\\
v_{2}%
\end{array}
\right)  \,\ .\label{vevscalarmssm}%
\end{equation}

\section{Mass Diagonalization}

\label{apend:massdiag}

The mass matrix of the ``up" quark sector ($Y_u$) (quark with charge
$+2/3$)  and of the charged leptons ($Y_l$) are written as the following: 
\begin{equation}
Y_{u}=\left(
\begin{array}
[c]{ccc}%
y_{11}^{u} & y_{12}^{u} & 0\\
y_{21}^{u} & y_{22}^{u} & 0\\
y_{31}^{u} & y_{32}^{u} & 0
\end{array}
\right)  \cdot v_{2}\,\ ,\,\ Y_{l}=\left(
\begin{array}
[c]{ccc}%
y_{11}^{l} & y_{12}^{l} & 0\\
y_{21}^{l} & y_{22}^{l} & 0\\
y_{31}^{l} & y_{32}^{l} & 0
\end{array}
\right)  \cdot v_{1}\,\ ,\label{umassmssm}%
\end{equation}
while to the case of ``down " quark sector  the mass matrix is given by
\begin{equation}
Y_{d}=\left(
\begin{array}
[c]{ccc}%
0 & 0 & y_{13}^{d}\\
0 & 0 & y_{23}^{d}\\
0 & 0 & y_{33}^{d}%
\end{array}
\right)  \cdot v_{1}\,\ ,\label{dmassmssm}%
\end{equation}
where $v_{1}$ and $v_{2}$ are VEVs of $H_{1}$ and $H_{2}$ respectively.

The fermion's mass matrix is diagonalized using two unitary matrices, $D$ and
$E$. Then we can write the diagonal mass matrix as
\begin{equation}
M_{diag}^{2}=DY_{f}^{T}\cdot Y_{f}D^{-1}=E^{\ast}Y_{f}\cdot Y_{f}^{T}(E^{\ast
})^{-1},\label{m2}
\end{equation}
where $f$ can represent any ``up", ``down" quarks or any charged lepton.

After the diagonalization we have defined the followings parameters
\begin{eqnarray}
t_{u}  & =&\left(  (y_{11}^{u})^{2}+(y_{12}^{u})^{2}+(y_{21}^{u})^{2}+
(y_{22}^{u})^{2}+(y_{31}^{u})^{2}+(y_{32}^{u})^{2}\right)  \cdot v_{2},\nonumber\\
t_{l}  & =&\left(  (y_{11}^{l})^{2}+(y_{12}^{l})^{2}+(y_{21}^{l})^{2}+(y_{22}^{l})^{2}+
(y_{31}^{l})^{2}+(y_{32}^{l})^{2}\right)  \cdot v_{1},\nonumber\\
r_{u}  & =&\sqrt{\left(  t_{u}\right)  ^{2}-4((y_{12}^{u})^{2}u_{u}-2y_{11}%
^{u}y_{12}^{u}v_{u}+x_{u}^{2}+(y_{11}^{u})^{2}z_{u})}\cdot v_{2},\nonumber\\
u_{u}  & =&(y_{21}^{u})^{2}+(y_{31}^{u})^{2},\,\ v_{u}=y_{31}^{u}y_{32}^{u}+y_{21}^{u}y_{22}^{u}
, \nonumber \\ 
x_{u}&=&y_{21}^{u}y_{22}^{u}-y_{31}^{u}y_{32}^{u},\,\ 
z_{u}=(y_{22}^{u})^{2}+(y_{32}^{u})^{2},\nonumber\\
r_{l}  & =&\sqrt{\left(  t_{l}\right)  ^{2}-4((y_{12}^{l})^{2}u_{l}-2
y_{11}^{l}y_{12}^{l}v_{l}+x_{l}^{2}+(y_{11}^{l})^{2}z_{l})}\cdot v_{1},\nonumber\\
u_{l}  & =&(y_{21}^{l})^{2}+(y_{31}^{l})^{2},\,\ 
v_{l}=y_{31}^{l}y_{32}^{l}+y_{21}^{l}y_{22}^{l}, \nonumber \\ 
x_{l}&=&y_{21}^{l}y_{22}^{l}-y_{31}^{l}y_{32}^{l}%
,\,\ z_{l}=(y_{22}^{l})^{2}+(y_{32}^{l})^{2},\nonumber\\
t_{d}  & =&\left(  (y_{13}^{d})^{2}+(y_{23}^{d})^{2}+(y_{33}^{d})^{2}\right)  \cdot
v_{1},\,\ r_{d}=t_{d}.
\label{autovalor3}
\end{eqnarray}

\section{Triplet Model (SUSYLRT)}

\label{apend:susylrt}

\begin{table}[t]
\center
\renewcommand{\arraystretch}{1.5}
\begin{tabular}
[c]{|l|cc|cc|}\hline
Superfield & Usual Particle & Spin & Superpartner & Spin\\\hline\hline
\quad$\hat{V}^{\prime}$ (U(1)) & $B_{m}$ & 1 & $\tilde{B}\,\,$ & $\frac{1}{2}%
$\\
\quad$\hat{V}^{i}_{L}$ ($SU(2)_{L}$) & $W^{i}_{mL}$ & 1 & $\tilde{W}_{L}^{i}$
& $\frac{1}{2}$\\
\quad$\hat{V}^{i}_{R}$ ($SU(2)_{R}$) & $W^{i}_{mR}$ & 1 & $\tilde{W}_{R}^{i}$
& $\frac{1}{2}$\\
\quad$\hat{V}^{a}_{c} (SU(3))$ & $g^{a}_{m}$ & 1 & $\tilde{ g}^{a}$ &
$\frac{1}{2}$\\\hline
\quad$\hat{Q}_{i}\sim({\bf3},{\bf2},{\bf1},1/3)$ & $(u_{i},\,d_{i})_{iL}$ &
$\frac{1}{2}$ & $(\tilde{ u}_{iL},\,\tilde{ d}_{iL})$ & 0\\
\quad$\hat{Q}^{c}_{i}\sim({\bf3^{*}},{\bf1},{\bf2},-1/3)$ & $(d^{c}_{i}%
,\,-u^{c}_{i})_{iL}$ & $\frac{1}{2}$ & $(\tilde{ d}^{c}_{iL},\,- \tilde{
u}^{c}_{iL})$ & 0\\\hline
\quad$\hat{L}_{a}\sim({\bf1},{\bf2},{\bf1},-1)$ & $(\nu_{a},\,l_{a})_{aL}$
& $\frac{1}{2}$ & $(\tilde{ \nu}_{aL},\,\tilde{ l}_{aL})$ & 0\\
\quad$\hat{L}^{c}_{a}\sim({\bf1},{\bf1},{\bf2},1)$ & $(l^{c}_{a},\,-
\nu^{c}_{a})_{aL}$ & $\frac{1}{2}$ & $(\tilde{ l}^{c}_{aL},\,- \tilde{ \nu
}^{c}_{aL})$ & 0\\\hline
\quad$\hat{\Delta}_{L}\sim({\bf1},{\bf3},{\bf1},2)$ & $\left(
\begin{array}
[c]{cc}%
\frac{\delta_{L}^{+}}{\sqrt{2}} & \delta_{L}^{++}\\
\delta_{L}^{0} & \frac{-\delta_{L}^{+}}{\sqrt{2}}%
\end{array}
\right) $ & 0 & $\left(
\begin{array}
[c]{cc}%
\frac{\tilde{\delta}_{L}^{+}}{\sqrt{2}} & \tilde{\delta}_{L}^{++}\\
\tilde{\delta}_{L}^{0} & \frac{-\tilde{\delta}_{L}^{+}}{\sqrt{2}}%
\end{array}
\right) $ & $\frac{1}{2}$\\
\quad$\hat{\Delta}^{\prime}_{L}\sim({\bf1},{\bf3},{\bf1},-2)$ & $\left(
\begin{array}
[c]{cc}%
\frac{\delta_{L}^{\prime-}}{\sqrt{2}} & \delta_{L}^{\prime0}\\
\delta_{L}^{\prime--} & \frac{-\delta_{L}^{\prime-}}{\sqrt{2}}%
\end{array}
\right) $ & 0 & $\left(
\begin{array}
[c]{cc}%
\frac{\tilde{\delta}_{L}^{\prime-}}{\sqrt{2}} & \tilde{\delta}_{L}^{\prime0}\\
\tilde{\delta}_{L}^{\prime--} & \frac{-\tilde{\delta}_{L}^{\prime-}}{\sqrt{2}}%
\end{array}
\right) $ & $\frac{1}{2}$\\
\quad$\hat{\delta}^{c}_{L}\sim({\bf1},{\bf1},{\bf3},-2)$ & $\left(
\begin{array}
[c]{cc}%
\frac{\lambda_{L}^{-}}{\sqrt{2}} & \lambda_{L}^{0}\\
\lambda_{L}^{--} & \frac{-\lambda_{L}^{-}}{\sqrt{2}}%
\end{array}
\right) $ & 0 & $\left(
\begin{array}
[c]{cc}%
\frac{\tilde{\lambda}_{L}^{-}}{\sqrt{2}} & \tilde{\lambda}_{L}^{0}\\
\tilde{\lambda}_{L}^{--} & \frac{-\tilde{\lambda}_{L}^{-}}{\sqrt{2}}%
\end{array}
\right) $ & $\frac{1}{2}$\\
\quad$\hat{\delta}^{\prime c}_{L}\sim({\bf1},{\bf1},{\bf3},2)$ & $\left(
\begin{array}
[c]{cc}%
\frac{\lambda_{L}^{\prime+}}{\sqrt{2}} & \lambda_{L}^{\prime++}\\
\lambda_{L}^{\prime0} & \frac{-\lambda_{L}^{\prime+}}{\sqrt{2}}%
\end{array}
\right) $ & 0 & $\left(
\begin{array}
[c]{cc}%
\frac{\tilde{\lambda}_{L}^{\prime+}}{\sqrt{2}} & \tilde{\lambda}_{L}%
^{\prime++}\\
\tilde{\lambda}_{L}^{\prime0} & \frac{-\tilde{\lambda}_{L}^{\prime+}}{\sqrt
{2}}%
\end{array}
\right) $ & $\frac{1}{2}$\\\hline
\quad$\hat{\Phi} \sim\left(  {\bf1},{\bf2},{\bf2},0\right) $ & $\left(
\begin{array}
[c]{cc}%
\phi_{1}^{0} & \phi_{1}^{+}\\
\phi_{2}^{-} & \phi_{2}^{0}%
\end{array}
\right) $ & 0 & $\left(
\begin{array}
[c]{cc}%
\tilde{\phi}_{1}^{0} & \tilde{\phi}_{1}^{+}\\
\tilde{\phi}_{2}^{-} & \tilde{\phi}_{2}^{0}%
\end{array}
\right) $ & $\frac{1}{2}$\\
\quad$\hat{\Phi}^{\prime} \sim\left(  {\bf1},{\bf2},{\bf2},0\right) $ & $\left(
\begin{array}
[c]{cc}%
\chi_{1}^{0} & \chi_{1}^{+}\\
\chi_{2}^{-} & \chi_{2}^{0}%
\end{array}
\right) $ & 0 & $\left(
\begin{array}
[c]{cc}%
\tilde{\chi}_{1}^{0} & \tilde{\chi}_{1}^{+}\\
\tilde{\chi}_{2}^{-} & \tilde{\chi}_{2}^{0}%
\end{array}
\right) $ & $\frac{1}{2}$\\\hline
\end{tabular}
\caption{Particle content of SUSYLRT.}
\label{tab:SUSYLRT}
\end{table}

The particle content of the model is given at Tab.(\ref{tab:SUSYLRT}) (for
recent work see for example \cite{phenosusylr} and references therein). In 
parentheses it appears the transformation properties under the respective 
$(SU(3)_{C},SU(2)_{L},SU(2)_{R},U(1)_{B-L})$. The
Lagrangian is given by:
\begin{equation}
{\cal L}_{SUSYLRT}={\cal L}_{Lepton}+{\cal L}_{Quarks}+{\cal L}_{Gauge}+{\cal L}_{Higgs},
\end{equation}
where
\begin{eqnarray}
{\cal L}_{Lepton}  & =& \int d^{4}\theta\;\left[  \,\hat{\bar{L}}%
_{aL}e^{2gT^{i}\hat{V}_{L}^{i}+g^{\prime}(-1)\hat{V}^{\prime}}\hat{L}%
_{aL}+\hat{\bar{L}}_{aL}^{c}e^{2gT^{i}\hat{V}_{R}^{i}+g^{\prime}(1)\hat
{V}^{\prime}}\hat{L}_{aL}^{c}\right]  ,\nonumber\label{susylrt term}\\
{\cal L}_{Quarks}  & =& \int d^{4}\theta\;\left[  \,
\hat{\bar{Q}}_{i L}e^{2g_{s}T^{a}\hat{V}_{c}^{a}+2gT^{i}\hat{V}_{L}^{i}+g^{\prime}\left(
\frac{1}{3}\right)  \hat{V}^{\prime}}\hat{Q}_{iL}\right.  \nonumber\\
& +&\left.  \hat{\bar{Q}}_{i L}^{c}e^{2g_{s}\bar{T}^{a}\hat{V}_{c}^{a}
+2gT^{i}\hat{V}_{R}^{i}+g^{\prime}\left(  \frac{-1}{3}\right)  \hat
{V}^{\prime}}\hat{Q}_{i L}^{c}\,\right]  ,\nonumber\\
{\cal L}_{Gauge}  & =& \frac{1}{4}\left\{  \int d^{2}\theta\;\left[
\sum_{a=1}^{8}W_{s}^{a\alpha}W_{s\alpha}^{a}+\sum_{i=1}^{3}W_{L}^{i\alpha
}W_{L\alpha}^{i}+\sum_{i=1}^{3}W_{R}^{i\alpha}W_{R\alpha}^{i}\right.  \right.
\nonumber\\
& +& \left.  \left.  W^{\prime\alpha}W_{\alpha}^{\prime}\right]  +h.c.\right\}
,
\end{eqnarray}
with $T^{i}=\tau^{i}/2$ is the generator of $SU(2)$ group while $T^{a}%
=\lambda^{a}/2$ is the generator of triplets of $SU(3)$ while $\bar{T}%
^{a}=\bar{\lambda}^{a}/2$ is the generator of the anti-triplet of $SU(3)$. We
make the usual assumption that the left and right couplings are equal,
$g_{L}=g_{R}=g$. The terms $W_{s}^{a\alpha},W_{L}^{i\alpha},W_{R}^{i\alpha}$
and $W^{\prime\alpha}$ are calculated using expressions analogous to that at
Eq.(\ref{fieldstrength}).

The last part of our Lagrangian reads:
\begin{eqnarray}
{\cal L}_{Higgs}  & =& \int d^{4}\theta\;Tr\left[  \,\hat{\bar{\Delta}}%
_{L}e^{2gT^{i}\hat{V}_{L}^{i}+g^{\prime}(2)\hat{V}^{\prime}}\hat{\Delta}%
_{L}+\hat{\bar{\Delta}}_{L}^{\prime}e^{2gT^{i}\hat{V}_{L}^{i}+g^{\prime
}(-2)\hat{V}^{\prime}}\hat{\Delta}_{L}^{\prime}\right.  \nonumber\\
& +&\left.  \hat{\bar{\delta}}_{L}^{c}e^{2gT^{i}\hat{V}_{R}^{i}+g^{\prime
}(-2)\hat{V}^{\prime}}\hat{\delta}_{L}^{c}+\hat{\bar{\delta}}_{L}^{\prime
c}e^{2gT^{i}\hat{V}_{R}^{i}+g^{\prime}(2)\hat{V}^{\prime}}\hat{\delta}%
_{L}^{\prime c}+\hat{\bar{\Phi}}e^{2gT^{i}\hat{V}_{L}^{i}+2gT^{i}\hat{V}%
_{R}^{i}}\hat{\Phi}\right.  \nonumber\\
& +&\left.  \hat{\bar{\Phi}}^{\prime}e^{2gT^{i}\hat{V}_{L}^{i}+2gT^{i}\hat
{V}_{R}^{i}}\hat{\Phi}^{\prime}\right]  +\int d^{2}\theta W+\int d^{2}%
\bar{\theta}\overline{W}.\label{The Supersymmetric Term prop 4}
\end{eqnarray}

The most general superpotential $W$ \cite{susylr} is given by
\begin{eqnarray}
W  & =&M_{\Delta}Tr(\hat{\Delta}_{L}\hat{\Delta}_{L}^{\prime})+M_{\delta^{c}%
}Tr(\hat{\delta}_{L}^{c}\hat{\delta}_{L}^{\prime c})+\mu_{1}Tr(\imath\tau
_{2}\hat{\Phi}\imath\tau_{2}\hat{\Phi})+\mu_{2}Tr(\imath\tau_{2}\hat{\Phi
}^{\prime}\imath\tau_{2}\hat{\Phi}^{\prime})\nonumber\\
& +&\mu_{3}Tr(\imath\tau_{2}\hat{\Phi}\imath\tau_{2}\hat{\Phi}^{\prime}%
)+f_{ab}Tr(\hat{L}_{a}\imath\tau_{2}\hat{\Delta}_{L}\hat{L}_{b})+f_{ab}%
^{c}Tr(\hat{L}_{a}^{c}\imath\tau_{2}\hat{\delta}_{L}^{c}\hat{L}_{b}%
^{c})\nonumber\\
& +&h_{ab}^{l}Tr(\hat{L}_{a}\hat{\Phi}\imath\tau_{2}\hat{L}_{b}^{c})+\tilde
{h}_{ab}^{l}Tr(\hat{L}_{a}\hat{\Phi}^{\prime}\imath\tau_{2}\hat{L}_{b}%
^{c})+h_{ij}^{q}Tr(\hat{Q}_{i}\hat{\Phi}\imath\tau_{2}\hat{Q}_{j}%
^{c})\nonumber\\
& +&\tilde{h}_{ij}^{q}Tr(\hat{Q}_{i}\hat{\Phi}^{\prime}\imath\tau_{2}\hat
{Q}_{j}^{c})+W_{NR}.\label{suplr}%
\end{eqnarray}
Where $h^{l},\tilde{h}^{l},h^{q}$ and $\tilde{h}^{q}$ are the Yukawa couplings
for the leptons and quarks, respectively, and $f$ and $f^{c}$ are the
couplings for the triplets scalar bosons. We must emphasize that due to the
conservation of $B-L$ symmetry, $\Delta_{L}^{\prime}$ and 
$\delta_{L}^{\prime c}$ do not couple with the leptons and quarks. Here $W_{NR}$ denotes (possible)
non-renormalizable terms arising from higher scale physics or Planck scale
effects \cite{Chacko:1997cm}. This model can be embedded in a supersymmetric
grand unified theory as $SO(10)$ \cite{moha}.

In addition, we have also to include soft supersymmetry breaking terms, they 
 are:
\begin{eqnarray}
{\cal L}_{soft}  & =&\left[  m_{L_{L}}^{2}\tilde{L}_{L}^{\dagger}\tilde
{L}_{L}+m_{L_{R}}^{2}\tilde{L}_{L}^{c\dagger}\tilde{L}_{L}^{c}+m_{Q_{L}}%
^{2}\tilde{Q}_{L}^{\dagger}\tilde{Q}_{L}+m_{Q_{R}}^{2}\tilde{Q}_{L}^{c\dagger
}\tilde{Q}_{L}^{c}+m_{\Phi\Phi}^{2}\Phi^{\dagger}\Phi \right. \nonumber \\ 
&+&\left. m_{\Phi\Phi^{\prime}}^{2}\Phi^{\dagger}\Phi^{\prime}
+  m_{\Phi^{\prime}\Phi^{\prime}}^{2}\Phi^{\prime\dagger}\Phi^{\prime
}\right] -\left[
M_{1}^{2}Tr(\Delta_{L}\Delta_{L}^{\prime})+M_{2}^{2}Tr(\delta_{L}^{c}%
\delta_{L}^{\prime c})+M_{3}^{2}\Phi\Phi\right.  \nonumber\\
& +&\left.  M_{4}^{2}\Phi\Phi^{\prime}+M_{5}^{2}\Phi^{\prime}\Phi^{\prime
}+h.c.\right]  -\left[  
A^{LL}Tr(\tilde{L}\tau_{2}\Delta_{L}\tilde{L})+
\tilde{A}^{LL}Tr(\tilde{L}^{c}\tau_{2}\delta_{L}^{c}\tilde{L}^{c})\right.
\nonumber\\
& +&\left.  A^{LR}Tr(\tilde{L}\Phi\imath\tau_{2}\tilde{L}^{c})+
\tilde{A}^{LR}Tr(\tilde{L}\Phi^{\prime}\imath\tau_{2}\tilde{L}^{c})+
A^{QQ}Tr(\tilde{Q}\Phi\imath\tau_{2}\tilde{Q}^{c})\right.  \nonumber\\
& +&\left.  \tilde{A}^{QQ}Tr(\tilde{Q}\Phi^{\prime}\imath\tau_{2}\tilde{Q}%
^{c})+h.c.\right]  -\frac{1}{2}\left(  \sum_{i=1}^{8}m_{\tilde{g}}\tilde
{g}^{i}\tilde{g}^{i}+\sum_{i=1}^{3}m_{L}\tilde{W}_{L}^{i}\tilde{W}_{L}%
^{i}\right.  \nonumber\\
& +&\left.  \sum_{i=1}^{3}m_{R}\tilde{W}_{R}^{i}\tilde{W}_{R}^{i}+m^{\prime
}\tilde{B}\tilde{B}+h.c.\right).
\end{eqnarray}

The vacuum expectations values are given by \cite{fin1}
\begin{eqnarray}
\left\langle \Phi\right\rangle  & =&\frac{1}{\sqrt{2}}\left(
\begin{array}
[c]{cc}%
k_{1} & 0\\
0 & k_{1}^{\prime}%
\end{array}
\right)  ;\,\ \left\langle \Phi^{\prime}\right\rangle =\frac{1}{\sqrt{2}%
}\left(
\begin{array}
[c]{cc}%
k_{2}^{\prime} & 0\\
0 & k_{2}%
\end{array}
\right)  ;\nonumber\label{vevsusylr}\\
\left\langle \Delta_{L}\right\rangle  & =&\frac{1}{\sqrt{2}}\left(
\begin{array}
[c]{cc}%
0 & 0\\
\upsilon_{L} & 0
\end{array}
\right)  ;\,\ ;\,\ \left\langle \Delta_{L}^{\prime}\right\rangle =\frac
{1}{\sqrt{2}}\left(
\begin{array}
[c]{cc}%
0 & \upsilon_{L}^{\prime}\\
0 & 0
\end{array}
\right)  ;\nonumber\\
\left\langle \delta_{L}^{c}\right\rangle  & =&\frac{1}{\sqrt{2}}\left(
\begin{array}
[c]{cc}%
0 & \upsilon_{R}\\
0 & 0
\end{array}
\right)  ;\,\ \left\langle \delta_{L}^{\prime c}\right\rangle =\frac{1}%
{\sqrt{2}}\left(
\begin{array}
[c]{cc}%
0 & 0\\
\upsilon_{R}^{\prime} & 0
\end{array}
\right)  .\nonumber\\
\end{eqnarray}

\section{Doublet Model (SUSYLRD)}

\label{apend:susylrd}

\begin{table}[t]
\center
\renewcommand{\arraystretch}{1.5}
\begin{tabular}
[c]{|l|cc|cc|}\hline
Superfield & Usual Particle & Spin & Superpartner & Spin\\\hline\hline
\quad$\hat{V}^{\prime}$ (U(1)) & $B_{m}$ & 1 & $\tilde{B}\,\,$ & $\frac{1}{2}%
$\\
\quad$\hat{V}^{i}_{L}$ ($SU(2)_{L}$) & $W^{i}_{mL}$ & 1 & $\tilde{W}_{L}^{i}$
& $\frac{1}{2}$\\
\quad$\hat{V}^{i}_{R}$ ($SU(2)_{R}$) & $W^{i}_{mR}$ & 1 & $\tilde{W}_{R}^{i}$
& $\frac{1}{2}$\\
\quad$\hat{V}^{a}_{c} (SU(3))$ & $g^{a}_{m}$ & 1 & $\tilde{ g}^{a}$ &
$\frac{1}{2}$\\\hline
\quad$\hat{Q}_{i}\sim({\bf3},{\bf2},{\bf1},1/3)$ & $(u_{i},\,d_{i})_{iL}$ &
$\frac{1}{2}$ & $(\tilde{ u}_{iL},\,\tilde{ d}_{iL})$ & 0\\
\quad$\hat{Q}^{c}_{i}\sim({\bf3^{*}},{\bf1},{\bf2},-1/3)$ & $(d^{c}_{i}%
,\,-u^{c}_{i})_{iL}$ & $\frac{1}{2}$ & $(\tilde{ d}^{c}_{iL},\,- \tilde{
u}^{c}_{iL})$ & 0\\\hline
\quad$\hat{L}_{a}\sim({\bf1},{\bf2},{\bf1},-1)$ & $(\nu_{a},\,l_{a})_{aL}$
& $\frac{1}{2}$ & $(\tilde{ \nu}_{aL},\,\tilde{ l}_{aL})$ & 0\\
\quad$\hat{L}^{c}_{a}\sim({\bf1},{\bf1},{\bf2},1)$ & $(l^{c}_{a},\,-
\nu^{c}_{a})_{aL}$ & $\frac{1}{2}$ & $(\tilde{ l}^{c}_{aL},\,- \tilde{ \nu
}^{c}_{aL})$ & 0\\\hline
\quad$\hat{\chi_{1L}} \sim\left(  {\bf1},{\bf2},{\bf1},1 \right) $ & $(\chi_{1L}^{+},
\,\ \chi_{1L}^{0})$ & 0 & $(\tilde{\chi}_{1L}^{+}, \,\ \tilde{\chi}_{1L}^{0})$
& $\frac{1}{2}$\\
\quad$\hat{\chi_{2L}} \sim\left(  {\bf1},{\bf2},{\bf1},-1 \right) $ & $(\chi_{2L}^{0},
\,\ \chi_{2L}^{-})$ & 0 & $(\tilde{\chi}_{2L}^{0}, \,\ \tilde{\chi}_{2L}^{-})$
& $\frac{1}{2}$\\
\quad$\hat{\chi^{c}_{3L}} \sim\left(  {\bf1},{\bf1},{\bf2},-1 \right) $ & $(\chi_{3L}^{0},
\,\ \chi_{3L}^{-})$ & 0 & $(\tilde{\chi}_{3L}^{0}, \,\ \tilde{\chi}_{3L}^{-})$
& $\frac{1}{2}$\\
\quad$\hat{\chi^{c}_{4L}} \sim\left(  {\bf1},{\bf1},{\bf2},1 \right) $ & $(\chi_{4L}^{+},
\,\ \chi_{4L}^{0})$ & 0 & $(\tilde{\chi}_{4L}^{+}, \,\ \tilde{\chi}_{4L}^{0})$
& $\frac{1}{2}$\\\hline
\quad$\hat{\Phi} \sim\left(  {\bf1},{\bf2},{\bf2},0\right) $ & $\left(
\begin{array}
[c]{cc}%
\phi_{1}^{0} & \phi_{1}^{+}\\
\phi_{2}^{-} & \phi_{2}^{0}%
\end{array}
\right) $ & 0 & $\left(
\begin{array}
[c]{cc}%
\tilde{\phi}_{1}^{0} & \tilde{\phi}_{1}^{+}\\
\tilde{\phi}_{2}^{-} & \tilde{\phi}_{2}^{0}%
\end{array}
\right) $ & $\frac{1}{2}$\\
\quad$\hat{\Phi}^{\prime} \sim\left(  {\bf1},{\bf2},{\bf2},0\right) $ & $\left(
\begin{array}
[c]{cc}%
\chi_{1}^{0} & \chi_{1}^{+}\\
\chi_{2}^{-} & \chi_{2}^{0}%
\end{array}
\right) $ & 0 & $\left(
\begin{array}
[c]{cc}%
\tilde{\chi}_{1}^{0} & \tilde{\chi}_{1}^{+}\\
\tilde{\chi}_{2}^{-} & \tilde{\chi}_{2}^{0}%
\end{array}
\right) $ & $\frac{1}{2}$\\\hline
\end{tabular}
\caption{Particle content of SUSYLRD.}
\label{tab:SUSYLRD}
\end{table}
This model contains the particle content given at
Tab.(\ref{tab:SUSYLRD}). 

The Lagrangian of this model is given by:
\begin{equation}
{\cal L}_{SUSYLRD}={\cal L}_{Lepton}+{\cal L}_{Quarks}+{\cal L}_{Gauge}+{\cal L}_{Higgs},
\end{equation}
where ${\cal L}_{Lepton},{\cal L}_{Quarks},{\cal L}_{Gauge}$ are
given by Eq.(\ref{susylrt term}). The last part of our Lagrangian is given
by:
\begin{eqnarray}
{\cal L}_{Higgs}  & =&\int d^{4}\theta\left[  \hat{\bar{\chi}}_{1}%
e^{2gT^{i}\hat{V}_{L}^{i}+g^{\prime}(2)\hat{V}^{\prime}}\hat{\chi}_{1}%
+\hat{\bar{\chi}}_{2}e^{2gT^{i}\hat{V}_{L}^{i}+g^{\prime}(-2)\hat{V}^{\prime}%
}\hat{\chi}_{2}+\hat{\bar{\chi}}_{3}^{c}e^{2gT^{i}\hat{V}_{R}^{i}+g^{\prime
}(-2)\hat{V}^{\prime}}\hat{\chi}_{3}^{c}\right.
\nonumber\label{The Supersymmetric Term prop 4}\\
& +&\left.  \hat{\bar{\chi}}_{4}^{c}e^{2gT^{i}\hat{V}_{R}^{i}+g^{\prime}%
(2)\hat{V}^{\prime}}\hat{\chi}_{4}^{c}+\hat{\bar{\Phi}}e^{2gT^{i}\hat{V}%
_{L}^{i}+2gT^{i}\hat{V}_{R}^{i}}\hat{\Phi}+\hat{\bar{\Phi}}^{\prime}%
e^{2gT^{i}\hat{V}_{L}^{i}+2gT^{i}\hat{V}_{R}^{i}}\hat{\Phi}^{\prime}\right]
\nonumber\\
& +&\int d^{2}\theta W+\int d^{2}\bar{\theta}\overline{W}.\nonumber\\
\end{eqnarray}

The most general superpotential and soft supersymmetry breaking Lagrangian for
this model are:
\begin{eqnarray}
W  & =&M_{\chi}\hat{\chi}_{1}\hat{\chi}_{2}+M_{\chi^{c}}\hat{\chi}_{3}^{c}%
\hat{\chi}_{4}^{c}+\mu_{1}Tr(\tau_{2}\hat{\Phi}\tau_{2}\hat{\Phi})+\mu
_{2}Tr(\tau_{2}\hat{\Phi}^{\prime}\tau_{2}\hat{\Phi}^{\prime})+\mu_{3}%
Tr(\tau_{2}\hat{\Phi}\tau_{2}\hat{\Phi}^{\prime})\nonumber\\
& +&h_{ab}^{l}Tr(\hat{L}_{a}\hat{\Phi}\imath\tau_{2}\hat{L}_{b}^{c})+\tilde
{h}_{ab}^{l}Tr(\hat{L}_{a}\hat{\Phi}^{\prime}\imath\tau_{2}\hat{L}_{b}%
^{c})+h_{ij}^{q}Tr(\hat{Q}_{i}\hat{\Phi}\imath\tau_{2}\hat{Q}_{j}^{c})
\nonumber \\
&+&\tilde{h}_{ij}^{q}Tr(\hat{Q}_{i}\hat{\Phi}^{\prime}\imath\tau_{2}\hat{Q}%
_{j}^{c})
+W_{NR}.\label{suplrd}
\end{eqnarray}
and the soft terms reads
\begin{eqnarray}
{\cal L}_{soft}  & =&-\left[  m_{L_{L}}^{2}\tilde{L}_{L}^{\dagger}\tilde
{L}_{L}+m_{L_{R}}^{2}\tilde{L}_{L}^{c\dagger}\tilde{L}_{L}^{c}+m_{Q_{L}}%
^{2}\tilde{Q}_{L}^{\dagger}\tilde{Q}_{L}+m_{Q_{R}}^{2}\tilde{Q}_{L}^{c\dagger
}\tilde{Q}_{L}^{c}+m_{\Phi\Phi}^{2}\Phi^{\dagger}\Phi \right. \nonumber \\ 
&+&\left. m_{\Phi\Phi^{\prime}}^{2}\Phi^{\dagger}\Phi^{\prime}
+  m_{\Phi^{\prime}\Phi^{\prime}}^{2}\Phi^{\prime\dagger}\Phi^{\prime
}\right]  -\left[  M_{1}^{2}\chi_{1}\chi_{2}+M_{2}^{2}\chi_{3}^{c}\chi_{4}%
^{c}+M_{3}^{2}\Phi\Phi+M_{4}^{2}\Phi\Phi^{\prime} \right. \nonumber \\
&+& \left. M_{5}^{2}\Phi^{\prime}
\Phi^{\prime}+h.c.\right]  
-\left[  A^{LR}Tr(\tilde{L}\Phi\imath\tau_{2}\tilde{L}^{c})+
\tilde{A}^{LR}Tr(\tilde{L}\Phi^{\prime}\imath\tau_{2}\tilde{L}^{c}) 
\right. \nonumber \\ &+& \left.
A^{QQ}Tr(\tilde{Q}\Phi\imath\tau_{2}\tilde{Q}^{c})+
\tilde{A}^{QQ}Tr(\tilde{Q}\Phi^{\prime
}\imath\tau_{2}\tilde{Q}^{c})
+  h.c.\right]  \nonumber\\
& -&\frac{1}{2}\left(  \sum_{i=1}^{8}m_{\tilde{g}}\tilde{g}^{i}\tilde{g}%
^{i}+\sum_{i=1}^{3}m_{L}\tilde{W}_{L}^{i}\tilde{W}_{L}^{i}+\sum_{i=1}^{3}%
m_{R}\tilde{W}_{R}^{i}\tilde{W}_{R}^{i}+m^{\prime}\tilde{B}\tilde
{B}+h.c.\right)  \nonumber\\
\end{eqnarray}

The vacuum expectation values of the new scalars are
\begin{eqnarray}
\left\langle \chi_{1L}\right\rangle  & =&\frac{1}{\sqrt{2}}\left(
\begin{array}
[c]{c}%
0\\
\upsilon_{L}%
\end{array}
\right)  \,\ ,\,\ \left\langle \chi_{2L}\right\rangle =\frac{1}{\sqrt{2}%
}\left(
\begin{array}
[c]{c}%
\upsilon_{L}^{\prime}\\
0
\end{array}
\right)  ,\nonumber\\
\left\langle \chi_{3L}^{c}\right\rangle  & =&\frac{1}{\sqrt{2}}\left(
\begin{array}
[c]{c}%
\upsilon_{R}\\
0
\end{array}
\right)  ,\,\ \left\langle \chi_{4L}^{c}\right\rangle =\frac{1}{\sqrt{2}%
}\left(
\begin{array}
[c]{c}%
0\\
\upsilon_{R}^{\prime}%
\end{array}
\right)  .\label{vevscalardoublet}%
\end{eqnarray}


\begin{thebibliography}{99}                                                                                               %
\bibitem {sg}S. L. Glashow, {\sl Nucl. Phys.}{\bf 22}, 579, (1961).
\newline S. Weinberg, {\sl Phys. Rev. Lett.}{\bf 19}, 1264, (1967).
\newline A. Salam in {\sl Elementary Particle Theory: Relativistic Groups
and Analyticity}, Nobel Symposium N8 (Alquivist and Wilksells, Stockolm,
1968).\newline S. L. Glashow, J.Iliopoulos and L.Maini, 
{\sl Phys. Rev.}{\bf D 2}, 1285, (1970).

\bibitem {Moha99}R. N. Mohapatra hep-ph/9911272; Z. Berezhiani hep-ph/9602325

\bibitem {banks}T. Banks, {\sl Nucl. Phys.}{\bf B303}, 172, (1988).

\bibitem {ma}E. Ma, {\sl Phys. Rev.}{\bf D39}, 1922, (1989).

\bibitem {Pokors+90}H.P. Niles, M. Olechowski and S. Pokorski,
{\sl Phys. Lett.} {\bf 248}, 378 (1990).

\bibitem {Barbieri+96}R. Barbieri, G Dvali and L. J. Hall, 
{\sl Phys. Lett}{\bf B377}, 76 (1996).

\bibitem {Berez+05}Z. G. Berezhiani, hep-ph/9312222, Z. Brezhiani and A.
Rossi, {\sl Nuc. Phys. Proc. Suppl.} {\bf 101}, 410 (2001) ;Z. G. Berezhiani and
F. Nesti, hep-ph/0510011.

\bibitem {CHPT99}U. van Kolck, {\sl Prog. in Part and Nucl Phys}.
{\bf 43}, 337, (1999); J. Gasser and H. Leutwyler, {\sl Ann. of
Phys.}{\bf 158}, 142, (1980); {\sl Nucl Phys}.{\bf B250}, 465,
(1985); V. Bernard, N. Kaiser, Ulf-G. Meissner, {\sl Int. Jour. of Mod
Phys}.{\bf E4}, 193, (1995);V. Bernard, T. S. H. Lee, Ulf-G. Meissner,
{\sl Phys.Rep}.{\bf 246}, 315, (1994).

\bibitem {PotNN}S. L. Zhu, C. M. Maekawa, B. R. Holstein, M. J. Ramsey-Musolf
and U. van Kolck, {\sl Nucl Phys}.{\bf A748}, 435, (2005); S. L. Zhu, C.
M. Maekawa,\ G. Sacco, B. R. Holstein, M. J. Ramsey-Musolf, {\sl Phys.Rev}.
{\bf D65}, 033001, (2002); S. L. Zhu, C. M. Maekawa, B. R. Holstein, M. J.
Ramsey-Musolf, {\sl Phys. Rev. Lett}.{\bf 87}, 201802, (2001); U. van
Kolck, {\sl Phys. Rev}.{\bf C49}, 2932, (1994); C. Ordonez, L. Ray and
U. van Kolck, {\sl Phys. Rev. Lett.}.{\bf 72}, 1982, (1994); S.
Weinberg, {\sl Phys.Lett.}{\bf B295}, 114, (1992).

\bibitem {FFnuc}C. M. Maekawa and U. van Kolck, {\sl Phys.Lett.}{\bf B478}, 
73, (2000);C. M. Maekawa, J. S. da Veiga and U. van Kolck,
{\sl Phys.Lett.}{\bf B488}, 167, (2000); C. M. Maekawa, {\sl AIP Conf. Proc.} 
{\bf 739}, 663, (2005);

\bibitem {mssm}H. E. Haber and G. L. Kane, {\sl Phys. Rep.}{\bf 117}, 75 (1985).

\bibitem {lepmass}J. C. Montero, V. Pleitez and M. C. Rodriguez, {\sl Phys. Rev.}
{\bf D65}, 095008 (2002), and references therein.

\bibitem {dress}M. Dress, R. M. Godbole and P. Royr, 
{\it Theory and Phenomenology of Sparticles}
1st edition, World Scientific Publishing Co. Pte. Ltd., SDingapore, (2004).

\bibitem {wb}J. Wess and J. Bagger, {\it Supersymmetry and Supergravity}
2nd edition, Princeton University Press, Princeton NJ, (1992).

\bibitem {pdg}S. Eidelman \textit{et al.}, {\sl Phys. Lett.} {\bf B592} (2004).

\bibitem {susylr}K. Huitu, J. Maalampi and M. Raidal, {\sl Nucl. Phys.}{\bf B420}, 
449 (1994); C.S.Aulakh,A.Melfo and G.Senjanovic,
{\sl Phys.Rev.}{\bf D57},4174 (1998); G. Barenboim and N. Rius,
{\sl Phys. Rev.}{\bf D58}, 065010, (1998); N. Setzer and S. Spinner,
{\sl Phys. Rev.} {\bf D71}, 115010 (2005).

\bibitem {doublet}K. S. Babu.B. Dutta and R.N. Mohapatra, {\sl Phys.Rev.}
{\bf D65}:016005, (2002).

\bibitem {mfrank}M. Frank, {\sl Phys. Lett.} {\bf 540},269 (2002).

\bibitem {simp}R. M. Francis, M. Frank and C. S. Kalman, {\sl Phys. Rev.}
{\bf D43}, 2369 (1991).

\bibitem {10}L. Girardello and M. T. Grisaru, {\sl Nucl. Phys.}
{\bf B194} (1982) 65.

\bibitem {phenosusylr}M.Frank,I.Turan and A.C. de O\~{n}a, \textsl{Phys. Rev.}
\textbf{D72}, 075008 (2005);M. Frank, K. Huitu, T. Ruppell, hep-ph/0508056; M.
Frank, I. Turan, \textsl{Phys. Rev.}\textbf{D72}, 035008 (2005).

\bibitem {Chacko:1997cm}Z.~Chacko and R.~N.~Mohapatra, {\sl Phys. Rev.}
{\bf D58}, 015003 (1998); B.~Dutta and R.~N.~Mohapatra, {\sl Phys. Rev.}
{\bf D59}, 015018 (1999).

\bibitem {moha}R. N. Mohapatra, hep-ph/9801235.

\bibitem {fin1}K. Huitu, P.N. Pandita and K. Puolam\"aki, {\sl Phys. Lett.}
{\bf B423}, 97 (1998).

\bibitem {susy331}J. C. Montero, V. Pleitez and M. C. Rodriguez, 
{\sl Phys. Rev.}{\bf D65}, 035006 (2002); J. C. Montero, V. Pleitez and M. C.
Rodriguez, {\sl Phys. Rev.}{\bf D70},075004 (2004).
\end{thebibliography}
\end{document}